\documentclass[lettersize,journal]{IEEEtran}
\usepackage{amsmath,amsfonts}
\usepackage{algorithmic}
\usepackage{algorithm}
\usepackage{array}
\usepackage[caption=false,font=normalsize,labelfont=sf,textfont=sf]{subfig}
\usepackage{textcomp}
\usepackage{stfloats}
\usepackage{url}
\usepackage{verbatim}
\usepackage{graphicx}
\usepackage{cite}
\usepackage{tabularx}

\usepackage[utf8]{inputenc}
\usepackage[english]{babel}
\usepackage[usenames,dvipsnames,svgnames,table]{xcolor}
\usepackage{amsthm,bm} 
\usepackage{tikz}

\usepackage{amssymb}
\usepackage{gensymb}
\usepackage{setspace}
\usepackage{graphics}
\usepackage{pdflscape}
\usepackage{afterpage}
\usepackage{capt-of}
\usepackage{float}
\floatstyle{plaintop}
\restylefloat{table}

\usepackage[usenames,dvipsnames,svgnames,table]{xcolor}
            
\usepackage{float}
\usepackage{enumitem}

\usepackage{geometry}
\title{Data-driven control of COVID-19 in buildings:\\a reinforcement-learning approach}
\author{Ashkan Haji Hosseinloo, Saleh Nabi, Anette Hosoi, and Munther A. Dahleh \IEEEmembership{Fellow, IEEE}
\thanks{A. Haji Hosseinloo is with Department of Electrical Engineering and Computer
Science, Massachusetts Institute of Technology, Cambridge, MA 02139 USA
(e-mail: ashkanhh@mit.edu).}
\thanks{S. Nabi is with Mitsubishi Electric Research Laboratories, Cambridge, MA 02139 USA
(e-mail: nabi@merl.com).}
\thanks{A. Hosoi is with Department of Mechanical Engineering, Massachusetts Institute of Technology, Cambridge, MA 02139 USA
(e-mail: peko@mit.edu).}
\thanks{M. A. Dahleh is with Department of Electrical Engineering and Computer
Science, Massachusetts Institute of Technology, Cambridge, MA 02139 USA
(e-mail: dahleh@mit.edu).}
}
\date{}

\begin{document}
\maketitle

\begin{abstract}
In addition to its public health crisis, COVID-19 pandemic has led to the shutdown and closure of workplaces with an estimated total cost of more than \$16 trillion. Given the long hours an average person spends in buildings and indoor environments, this research article proposes data-driven control strategies to design optimal indoor airflow to minimize the exposure of occupants to viral pathogens in built environments. A general control framework is put forward for designing an optimal velocity field and proximal policy optimization, a reinforcement learning algorithm is employed to solve the control problem in a data-driven fashion. The same framework is used for optimal placement of disinfectants to neutralize the viral pathogens as an alternative to the airflow design when the latter is practically infeasible or hard to implement. We show, via simulation experiments, that the control agent learns the optimal policy in both scenarios within a reasonable time. The proposed data-driven control framework in this study will have significant societal and economic benefits by setting the foundation for an improved methodology in designing case-specific infection control guidelines that can be realized by affordable ventilation devices and disinfectants.
%
\end{abstract}
\def\abstractname{Note to Practitioners}
\begin{abstract}
This paper is motivated by the problem of COVID-19 infection spread in enclosed spaces but it also applies to other airborne pathogens. Airborne disease contagion often takes place in indoor environments; however, ventilation systems are almost never designed to take this into account so as to contain the spread of the pathogens. This is mainly because airflow design requires solving high-dimensional nonlinear partial differential equations known as Navier Stokes equations in fluid dynamics. In this paper, we propose a data-driven approach for solving the control problem of pathogen containment without solving the fluid dynamics equations. To this end, we first mathematically formulate the problem as an optimal control problem and then cast it as a reinforcement learning (RL) task. Reinforcement learning is the data-driven science of sequential decision-making and control in which the controller finds an optimal solution by systematic trial and error and without access to the system dynamics, i.e. fluid and pathogen dynamics in this paper. We employ an state-of-the-art RL algorithm, called PPO, to solve for optimal airflow in a room so as to minimize the exposure risk of occupants. Once it is calculated, the optimal airflow could be realized, via reverse engineering, by proper placement of the ventilation equipment, e.g. inlets, outlets, and fans. As an alternative to the airflow design, we use the same proposed data-driven techniques to find an optimal placement for pathogen disinfectants if there exists one, such as, hydrogen peroxide for COVID-19. Our results show the efficacy of our data-driven approach in designing an steady-state controller with full access to the system states. In future research, we will address the controller design with sparse measurements of the system states.
%
\end{abstract}
\section{Introduction}
In addition to its public health crisis, COVID-19 pandemic led to the shutdown and closure of workplaces, retail and commercial spaces, schools, and restaurants among many others. The lockdown has severely impacted the US economy and caused millions of temporary and permanent job losses in the US alone. The US unemployment rose higher in the first three months of COVID-19 than it did in two years of the Great Recession: 14.4\% in April 2020 versus  10.6\% in January 2010. Safe reopening and containing the spread of COVID-19 in indoor spaces is an important step towards economy recovery without risking people's health. This requires a good understanding of the disease transmission and designing effective engineering controls that is the purpose of this research article.

Although WHO and CDC ignored, in the beginning, the importance of airborne transmission for the disease, observational studies and computational models \cite{li2020evidence,hamner2020high,zhang2020identifying, kohanski2020review, morawska2020time} show that COVID-19 can remain aloft in air for a few hours and pose a risk of exposure at distances beyond the commonly adopted 6-feet social distancing. Hence, indoor ventilation and airflow play a big role in containment or spread of COVID-19 pathogens, especially knowing that people in industrialized countries spend more than 90\% of their lifetime indoors \cite{zhang2005indoor}. This is not specific to COVID-19 and the ventilation potential for preventing airborne disease transmission has been highlighted in the past \cite{li2007role, aliabadi2011preventing, qian2018ventilation}. Despite their importance, indoor ventilation and airflow are usually not designed for disease preventive purposes. Negative pressurized isolation rooms for patients with airborne diseases in hospitals are the only widely-adopted use of airflow design for preventive purposes. Personalized ventilation (PV) which delivers fresh air directly to the occupant's breathing zone is another design concept that can be leveraged for airborne infection control \cite{nielsen2009control, pantelic2009personalized, habchi2016ceiling, xu2018personalized, ding2020hvac}. PVs are not well explored and can be costly to design and implement.

Designing an effective preventive airflow requires a good transmission model for the disease. Most infection control strategies are mainly based on overly-simplified models of disease transmission developed in the 1930s \cite{bourouiba2020turbulent} which can limit the effectiveness of the resulting guidelines. The COVID-19 pandemic, however, gave rise to many studies exploring the host-to-host transmission of the disease with a wide spectrum of model complexity, from analytical well-mixed models to fully-blown 3D Navier-Stokes simulations. Burridge et. al. \cite{Burridge2021Predictive} and Luhar \cite{luhar2020airborne} used a well-mixed model to calculate the pathogen concentration and assess the infection risk in built environments. Balachandar et. al.\cite{balachandar2020host} developed a simple model for the time evolution of droplet/aerosol concentration based on a theoretical analysis  of the relevant physical processes. Their model ignores ambient mean flow and, hence, is not suitable for ventilation and airflow design. On the other end of the spectrum of the model complexity, computational fluid dynamics (CFD) simulations were adopted to solve Navier-Stokes equations coupled with pathogen transport equation to solve for spatiotemporal pathogen concentration in supermarkets \cite{vuorinen2020modelling}, urban buses \cite{mesgarpour2021prediction, zhang2021disease}, and a music classroom \cite{narayanan2021airborne}. Such detailed models are computationally too expensive for optimization and controller design. This is even more problematic for online control and optimization where computational burden can easily make the real-time design impossible. In the middle range of the model complexity, Lau et. el. \cite{lau2020modelling, lau2020predicting} modeled pathogen concentration evolution as an advection-diffusion equation with uniform velocity field. Using the concentration, different risk measures, such as, time to infection were calculated.

As discussed above, transmission models of airborne diseases are either too simplistic for an effective controller design or too complex for a computationally-feasible model-based controller design. Furthermore, much is still unknown about bio- and fluid-physics of COVID-19 pathogens, and hence, physics-based models may ignore some important aspects of the transmission dynamics. Also, in addition to the complex and not-fully-understood transmission dynamics, airflow design depends very much on the interior layout of the space that is often subject to continuous change, e.g., by changing seating layout in a restaurant or classroom. This warrants building a new model for the space and redesigning its controller every time there is a change in the space layout. For the above-mentioned reasons, a model-free and data-driven approach for airflow design is a better alternative. That is why we take a data-driven approach, namely, reinforcement learning for the controller design in this article.

The efforts for applying reinforcement learning (RL), and deep reinforcement learning (DRL) to fluid mechanics started only a few years ago in 2016 \cite{reddy2016learning, gazzola2016learning} and are still at an early stage, with only a handful of pioneering studies. In terms of the specific applications within the fluid mechanics, majority of these studies focus on drag reduction on a two-dimensional cylinder submerged in a fluid flow \cite{rabault2019artificial, fan2020reinforcement, tang2020robust, paris2021robust, ren2021applying}. For instance in \cite{rabault2019artificial} DRL with proximal policy optimization (PPO) algorithm is employed to reduce the drag by 8\% via controlling mass flow rates of two small jets on the sides of the cylinder. The second most-explored fluid mechanics application is fish swimming \cite{gazzola2016learning, novati2017synchronisation, verma2018efficient, yan2020numerical, zhu2021numerical}. For example in \cite{verma2018efficient}, the authors study the collective swimming of fishes and use DRL to find their optimal swimming strategy which turns out to be placing themselves in appropriate locations in the wake of other swimmers and intercepting judiciously their shed vortices.  

In this study, we employ data-driven and RL techniques to design effective indoor airflow in order to reduce the disease exposure risk for the occupants. We also apply the said techniques to optimally place pathogen neutralizers (disinfectants) in a room for a given airflow. To the best of our knowledge, this is the first application of RL in airborne disease transmission control. To this end, we first formulate the control problem in section \ref{Section:Control} after which the RL framework and methodology are discussed in section \ref{Section:RL}. Then we present and discuss the results in section \ref{Section:Results} before we conclude the paper with some remarks and future directions in section \ref{Section:Conclusion}. 
%
%
\section{Control problem formulation}\label{Section:Control}
\begin{figure}
 \centering
 \includegraphics[width = 1 \linewidth]{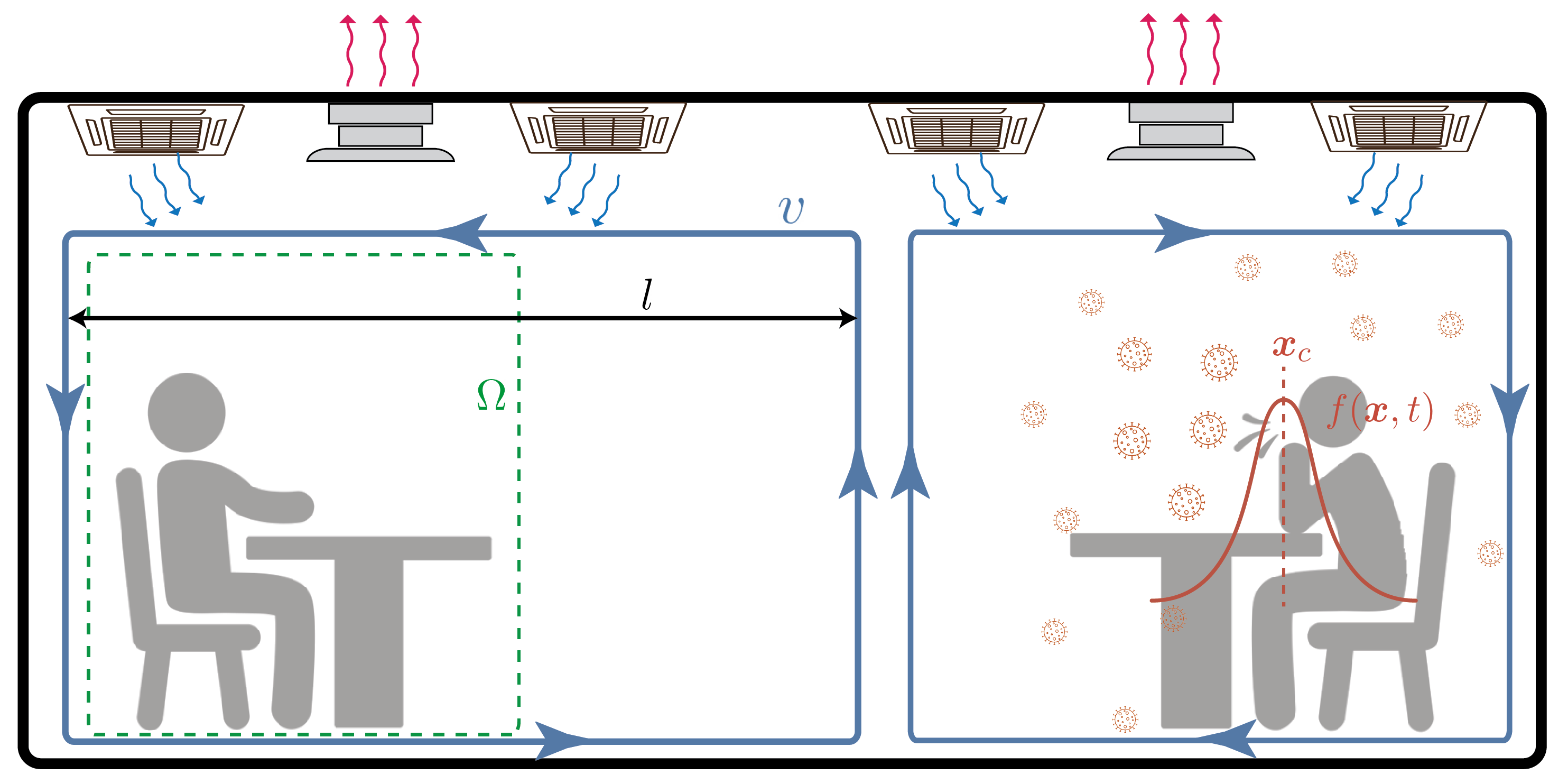}
 \caption{Control Problem 1: schematics of a room with the pathogen source, $f(\bm{x},t)$ (with a Gaussian spatial distribution centered at $\bm{x}_c$) and the region of interest, $\Omega$. A parameterized family of velocity field, namely, the double-vortex airflow is chosen in this study and is schematically shown by blue rectangles with arrows. The length of the left vortex is designated by $l$. }
 \label{Fig:P1Schematics}
\end{figure}

Let us consider, with no loss of generality, an exemplary case of a restaurant where airborne pathogens are released near a table with one or more infected customers at that table. We would like to design an airflow using available heating, ventilation, and air-conditioning (HVAC) system (e.g., fans and air-conditioners) that minimizes the exposure of the rest of the customers to the pathogens. This is the first of the two control problems we study in this paper and is schematically shown in Fig.\ref{Fig:P1Schematics}. Here, we consider the velocity field, $\bm{v}$, as the control variable and model transport dynamics of the virus by an advection-diffusion equation. We model the notion of exposure risk, that is also the control problem's performance metric $J_e$, as integral of the pathogen concentration, $c$, over a given time period, $[0, T]$, and a region of interest, $\Omega$. We then formulate the control problem as:
\begin{equation}
\begin{aligned}
\min_{\bm{v}} \quad & J_e = \int_{\Omega}\int_0^T c(\bm{x},t) \,dt\, d\bm{x}\\
\textrm{s.t.} \quad & \nabla.\bm{v}=0\\
  & \frac{\partial c}{\partial t}+\bm{v}.\nabla c - K\nabla^2 c=f(\bm{x},t)-\lambda c,
\end{aligned}
\label{Eq:ControlP1}
\end{equation}
where, the first and second constraints are incompressibility condition and the pathogen transport dynamics, respectively. Diffusion coefficient is denoted by $K$. Similar to \cite{lau2020modelling}, the overall effect of air and virus removal via the HVAC system is modeled by the term, $-\lambda c$. The coefficient $\lambda$ defines strength of the HVAC system and can be a function of time and space. Spatial coordinate and time are denoted by $\bm{x}$ and $t$, respectively, and $f(\bm{x},t)$ is the virus source. We assume Neumann boundary condition for the concentration, i.e. $\partial c/\partial \bm{n}=0$ where $\bm{n}$ is the normal to the boundary, $\partial \Omega$.
It is worth to mention that the actual control variables, in practice, are usually the location of the air inlet and outlet, as well as features of the inlet air, such as, temperature and velocity. Optimal values for these control variables can be found once we design the optimal velocity field, though it is not a trivial problem. Alternatively, the control problem could be formulated such that the above-mentioned variables are set as the control variables. In this case, airflow dynamics should be added to the problem constraints, e.g. in the form of Navier Stokes equations. This is beyond the scope of this study due to its computational complexity; however, the proposed control and RL framework in this study will directly apply to this formulation as well.

\begin{figure}
 \centering
 \includegraphics[width = 1 \linewidth]{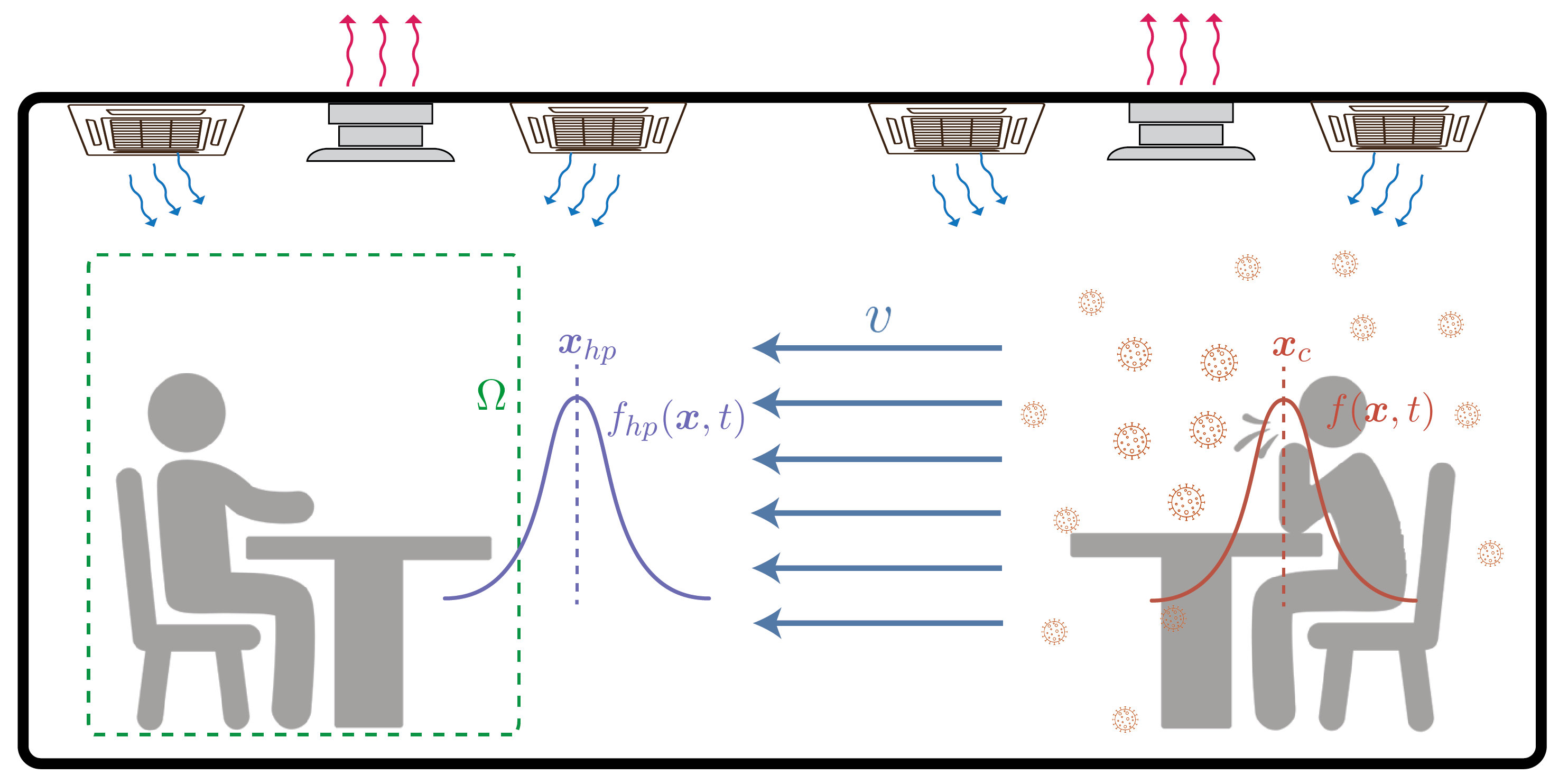}
 \caption{Control Problem 2: Schematics of a room with the pathogen and HP sources, $f(\bm{x},t)$ and $f_{hp}(\bm{x},t)$, both with Gaussian spatial distributions centered at $\bm{x}_c$ and $\bm{x}_{hp}$, respectively. The region of interest is denoted by $\Omega$. For this control problem, a constant uniform velocity field is considered in this study. The control objective is to find an optimal center position ($\bm{x}_{hp}$) for the HP disinfectant source.}
 \label{Fig:P2Schematics}
\end{figure}

In addition to the airflow design, an alternative solution to control the virus spread is to neutralize them. Hydrogen Peroxide (HP), $\textrm{H}_2 \textrm{O}_2$, particularly in its ionized state, has been shown to be an effective disinfectant for the COVID-19 virus \cite{cheng2020disinfection, schwartz2020decontamination, kenney2022hydrogen}. As the second control problem, we consider here another set-up where HP is used to neutralize and disinfect the COVID-19 pathogens (see Fig.\ref{Fig:P2Schematics}). The general set-up is as the first one, with the difference that we assume a fixed uniform airflow and try to optimize the location of the HP source so as to minimize the same performance metric in Eq.\ref{Eq:ControlP1}. The control problem is formulated as below:
%
%
%
\begin{equation}
\begin{aligned}
\min_{ \bm{x}_{hp}} \quad & J_e = \int_{\Omega}\int_0^T c(\bm{x},t) \,dt\, d\Omega\\
\textrm{s.t.} \quad & \nabla . \bm{v}=0\\
  & \frac{\partial c}{\partial t}+ \bm{v}.\nabla c - K \nabla^2 c =f(\bm{x},t)-\lambda c \, + g_1(c,c_{hp})\\
  & \begin{aligned}
  \frac{\partial c_{hp}}{\partial t}+ \bm{v}.\nabla c_{hp} - K_{hp} \nabla^2 c_{hp} =&f_{hp}(\bm{x},t)-\lambda c_{hp} \\
  &+ g_2(c,c_{hp}),
\end{aligned}
\end{aligned}
\label{Eq:ControlP2}
\end{equation}
%
%
%
where, $c_{hp}$, $K_{hp}$, and $f_{hp}(\bm{x},t)$ designate the concentration, diffusivity, and the source of HP, respectively. The chemical interaction between COVID-19 and HP particles are captured by the functions $g_1$ and $g_2$. Here, we consider a simple proportional model for these interactions as: $g_1(c,c_{hp}) = \alpha_1\, c\, c_{hp}$ and  $g_2(c,c_{hp}) = \alpha_2\, c\, c_{hp}$, where $\alpha_1$ and $\alpha_2$ are constants. We model the HP source, as well as, the COVID-19 source in both control problems (Eqs. \ref{Eq:ControlP1} and \ref{Eq:ControlP2}) as time-invariant, spatially Gaussian-distributed functions:
$R_{(.)}/\pi \epsilon \exp(-(\bm{x}-\bm{x}_{(.)})^2/\epsilon)$, where, the subscript $(.) = c \textrm{\,\, or\,\,} hp$, shows whether the parameter pertains to COVID or HP. The strength and spread of the source are decided by the parameters $R$ and $\epsilon$, respectively. We would like to emphasize that the decision variable in the second control problem defined in Eq.\ref{Eq:ControlP2} is the center location of the HP source, i.e., $\bm{x}_{hp}$.
%


%
\section{Reinforcement learning framework}\label{Section:RL}
Reinforcement Learning is the data-driven science of sequential decision making. It is about learning the optimal behavior/decisions in an environment to maximize a notion of cumulative or average reward. This optimal behavior is learned through interactions with the, often unknown, environment, similar to children exploring the world around them and learning the actions that help them achieve a goal.

In the RL framework, the agent, aka the controller, takes action $a$ at state $s$ and observes an immediate reward $r$ after moving to next state $s'$. The goal of the agent is to find an optimal policy $\pi^*(s)$, i.e., optimal control law that maximizes the discounted cumulative rewards in expectation, usually referred to as the value function $V^\pi(s)$. In practice, we often maximize the value function weighted by the state distribution under the policy $\rho^\pi(s)$, i.e., $\int_s{\rho(s)V(s)\, ds}$. For the control problem defined by Eq.\ref{Eq:ControlP1}, the state is the spatially-continuous concentration field and the action is the continuous velocity field. We define the immediate reward as:
\begin{equation}
r = \int_{t_1}^{t_2}\int_{\Omega} c(\bm{x},t)\, d\bm{x}\,dt,
\label{Eq:reward}
\end{equation}
where, $t_1$ and $t_2$ are timestamps at states $s_1$ and $s_2$, respectively. In the second control problem defined by Eq.\ref{Eq:ControlP2}, the state is the aggregation of both COVID-19 and HP concentration fields, as well as, the velocity field. The action is the center location of the HP source, i.e. $\bm{x}_{hp}$. The reward function remains the same as that in Eq.\ref{Eq:reward}.

Reinforcement learning algorithms can, in general, be categorized into policy-based and value-based methods. In a value-based method, the algorithm learns an estimate of the optimal value function. Q-learning is probably the most well-known and one of the earliest value-based RL methods\footnote{Q-learning is also used extensively as part of many policy-based algorithms to learn the value or action-value function that serves as the critic for the actor.}. The policy here is implicit and can be derived directly from the value function.

In policy-based methods, however, we explicitly build a representation of a policy which we improve by, e.g., a policy-gradient (PG) technique. Policy-based methods have a few advantages over their value-based counterparts. They can solve for both deterministic and stochastic optimal policies. Furthermore, in many cases the optimal policy has a simpler form than the optimal value function, and hence, it is easier to directly learn the optimal policy\cite{hosseinloo2020data, hosseinloo2020event}. They also handle continuous action space better than their value-based counterparts \cite{hosseinloo2022deterministic}.

Among different policy gradient algorithms, we use PPO \cite{schulman2017proximal}, one of the state-of-the-art algorithms for training our RL agent in this study. Compared to many other policy gradient algorithms, such as, trust region policy optimization (TRPO), PPO is mathematically less complex, and hence, computationally faster \cite{rabault2019artificial}. It is also shown to be often more data-efficient. The most important aspect of the PPO algorithm is its clipped surrogate objective which prevents taking big steps in potentially wrong directions when updating the policy using the gradient decent. In the next section, we delineate the simulation experiments, discuss how the PPO algorithm is applied to the two control problems, and present and discuss the results.
\section{Results and discussion}\label{Section:Results}
We use a Python library called FEniCSx to solve the advection-diffusion dynamics in Eqs.\ref{Eq:ControlP1} and \ref{Eq:ControlP2}. FEniCSx is a popular open-source computing platform for solving partial differential equations (PDEs) that is based on finite element (FE) methods \cite{alnaes2015fenics}. As a FE-based solver, FEniCSx requires PDEs in variational form, aka the weak form. The variational reformulation of the boundary-value problems in Eqs.\ref{Eq:ControlP1} and \ref{Eq:ControlP2} are presented in the appendix. A linear Lagrange element is used for both test and trial functions in the variational formulation.
\begin{table*}
    \begin{center}
        \begin{tabularx}{0.7\textwidth}{lcr}
        \firsthline
        parameter description                       & symbol            & numerical value \\
        \hline
        room length                                 & $l_x$             & 8 $m$ \\
        room height                                 & $l_y$             & 4 $m$  \\
        diffusion coeff. for COVID-19               & $K$               & 0.022 $m^2/s$ \\
        diffusion coeff. for HP                     & $K_{hp}$          & 0.022 $m^2/s$ \\
        HVAC air-exchange coeff. (COVID-19)         & $\lambda$         & 0.0085 $1/s$ \\
        HVAC air-exchange coeff. (HP)               & $\lambda_{hp}$    & 0.0085 $1/s$ \\
        source intensity for COVID-19               & $R$               & 2.5 $\textrm{particle}/s$ \\
        source intensity for HP                     & $R_{hp}$          & 2.5 $\textrm{particle}/s$ \\
        interaction coeff. between COVID-19 and HP  & $\alpha_1$        & 0.2 $m^2/\textrm{particle.\,} s$ \\
        interaction coeff. between HP and COVID-19  & $\alpha_2$        & 0.2 $m^2/\textrm{particle.\,} s$ \\
        \lasthline
        \end{tabularx}
        \caption{Parameters description for the advection-diffusion dynamics}
        \label{Table:params}
    \end{center}
\end{table*}
%
%
%
As discussed in section \ref{Section:RL}, PPO is a policy-gradient method for which we need a parameterized policy. For the control problem in Eq.\ref{Eq:ControlP1}, the policy to optimize is the velocity field $\bm{v}$ which we parameterize by the parameter vector $\bm{\theta}$ and denote it as $\bm{v}^{\bm{\theta}}$. For this control problem, we focus on a specific family of velocity field, often known as double-vortex velocity (schematically shown in Fig.\ref{Fig:P1Schematics}). We chose this particular family of velocity field for a number of reasons. First, it is relatively easy to realize this velocity field, e.g., by adjusting the location of inlets and outlets or the direction of the inflow air. Second, it can significantly reduce the dimension of the action space, or equivalently the size of the parameter vector $\bm{\theta}$, by expressing the vortex dynamics using its geometric center (or vortex length) and strength. In our case we consider the former only (fixing the strength) resulting in $\bm{\theta}=\{l\}$, where $l$ is the length of the left vortex. And last but not least, stemming from domain knowledge, the double- and multiple-vortex fields can potentially create a notion of \textit{air distancing} by isolating the infected region from a non-infected region in the room quite easily. The vortices in double-vortex flow are reminiscent of convection rolls encountered in forced convection in buildings \cite{chen1998zero, farahmand2017deep}. The two-dimensional double-vortex velocity field $\bm{v}^{\bm{\theta}}=\bm{v}^l$ is mathematically formulated as below:
%
%
\begin{equation} \label{Eq:DVortex}
  \bm{v}^{l} =
    \begin{cases}
      \bigl(w_x^l \sin(\frac{\pi x}{l})\cos(\frac{\pi y}{l_y}),\\ 
      -w^l_y\cos(\frac{\pi x}{l})\sin(\frac{\pi y}{l_y})\bigr): & \text{if } x\leq l\\[6pt]
      
      \bigl(w_x^r \sin(\frac{\pi (x-l)}{l_x-l})\cos(\frac{\pi y}{l_y}),\\ 
      -w^r_y\cos(\frac{\pi (x-l)}{l_x-l})\sin(\frac{\pi y}{l_y})\bigr): & \text{if } x>l,
    \end{cases}       
\end{equation}
where, $x$ and $y$ are the spatial coordinates in the horizontal and vertical directions, respectively, with $l_x$ and $l_y$ denoting the room length in their respective directions. The parameters $w$'s define the strength of the velocity field. These parameters cannot be chosen independently as the velocity field needs to satisfy the incompressibility condition ($\nabla.\bm{v}^l=0$). Enforcing this condition yields:
\begin{align} 
& w_x^l/l =  w_y^l/l_y \nonumber \\ 
& w_x^r/(l_x-l) = w_y^r/ly.
\end{align}
For all the simulations in this study we use $w_x^l=w_x^r =1.0\, m/s$. For the PPO implementation we use the open-source Stable Baselines3 (SB3) Python library. Stable Baselines3 is a set of reliable implementations of reinforcement learning algorithms in PyTorch \cite{stable-baselines3}. For user-defined environments SB3 requires a gym-compatible environment which we create using the FEniCSx library in Python.

\begin{figure}
 \centering
 \includegraphics[width = 1 \linewidth]{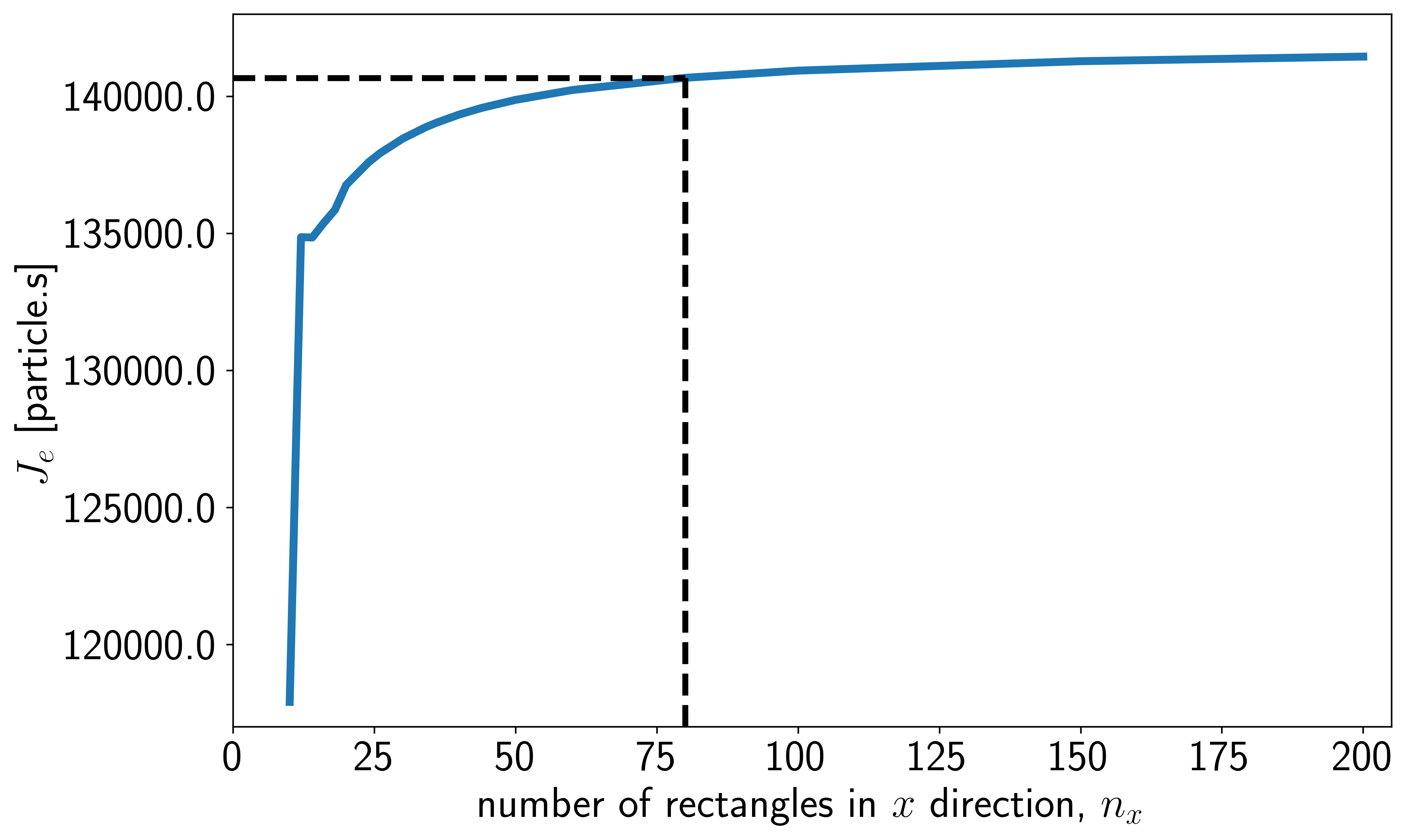}
 \caption{The performance metric, $J_e$ as a function of the mesh size, $n_x$ with parameters in Table \ref{Table:params}, left vortex length $l= 4.0\,m$, total simulation time $T=600\,s$, and the region of interest $\Omega=\text{entire room}$. $n_x=2n_y=80$ is chosen for all the subsequent simulations.}
 \label{Fig:MeshStudy}
\end{figure}

\begin{figure}
 \centering
 \includegraphics[width = 1 \linewidth]{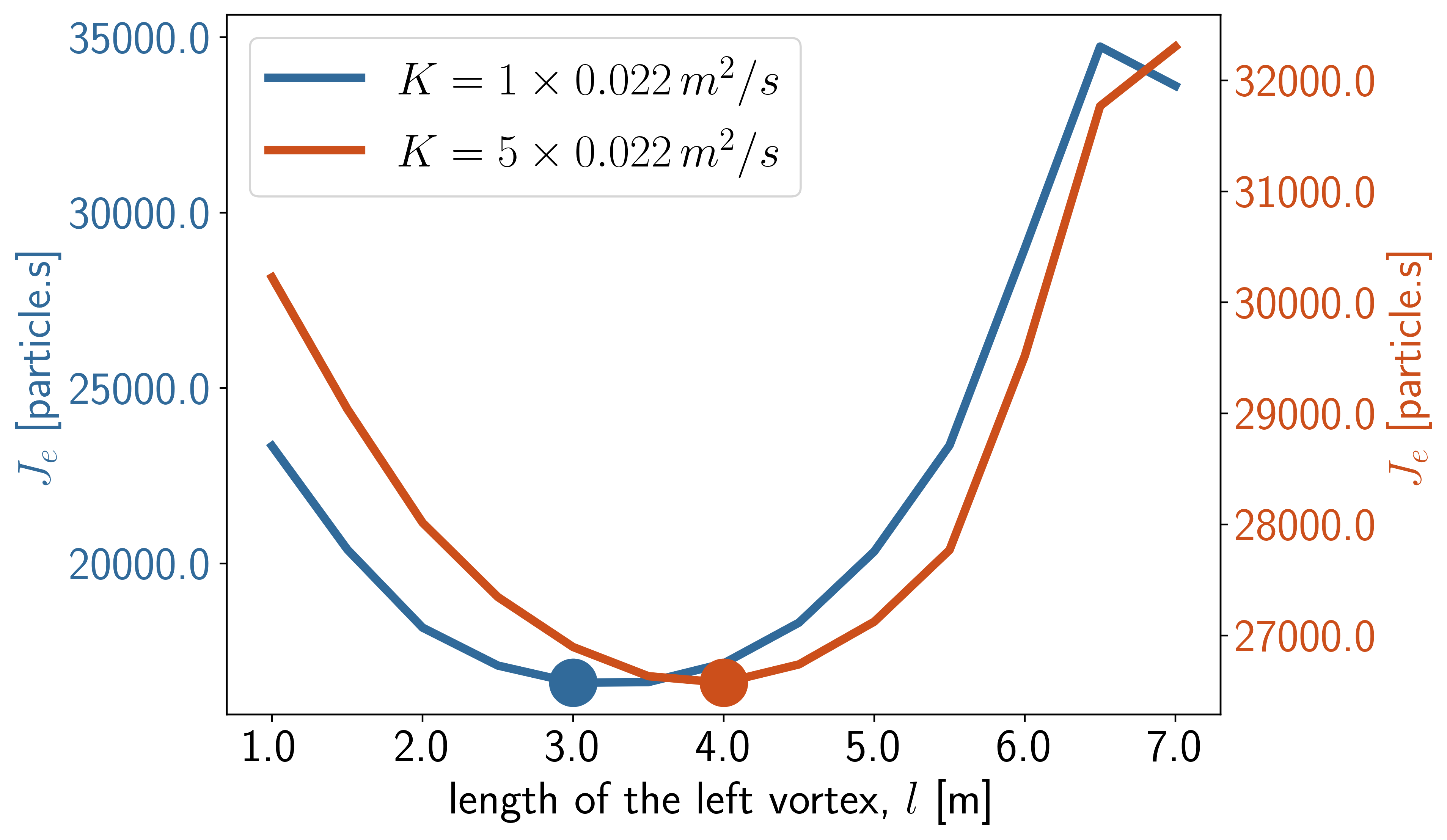}
 \caption{The performance metric, $J_e$ as a function of the left vortex length for two different values of diffusion coefficients. The optimal values minimizing the performance metric are marked by solid circles.}
 \label{Fig:DVortex_GroundTruth}
\end{figure}

\begin{figure}
 \centering
 \includegraphics[width = 1 \linewidth]{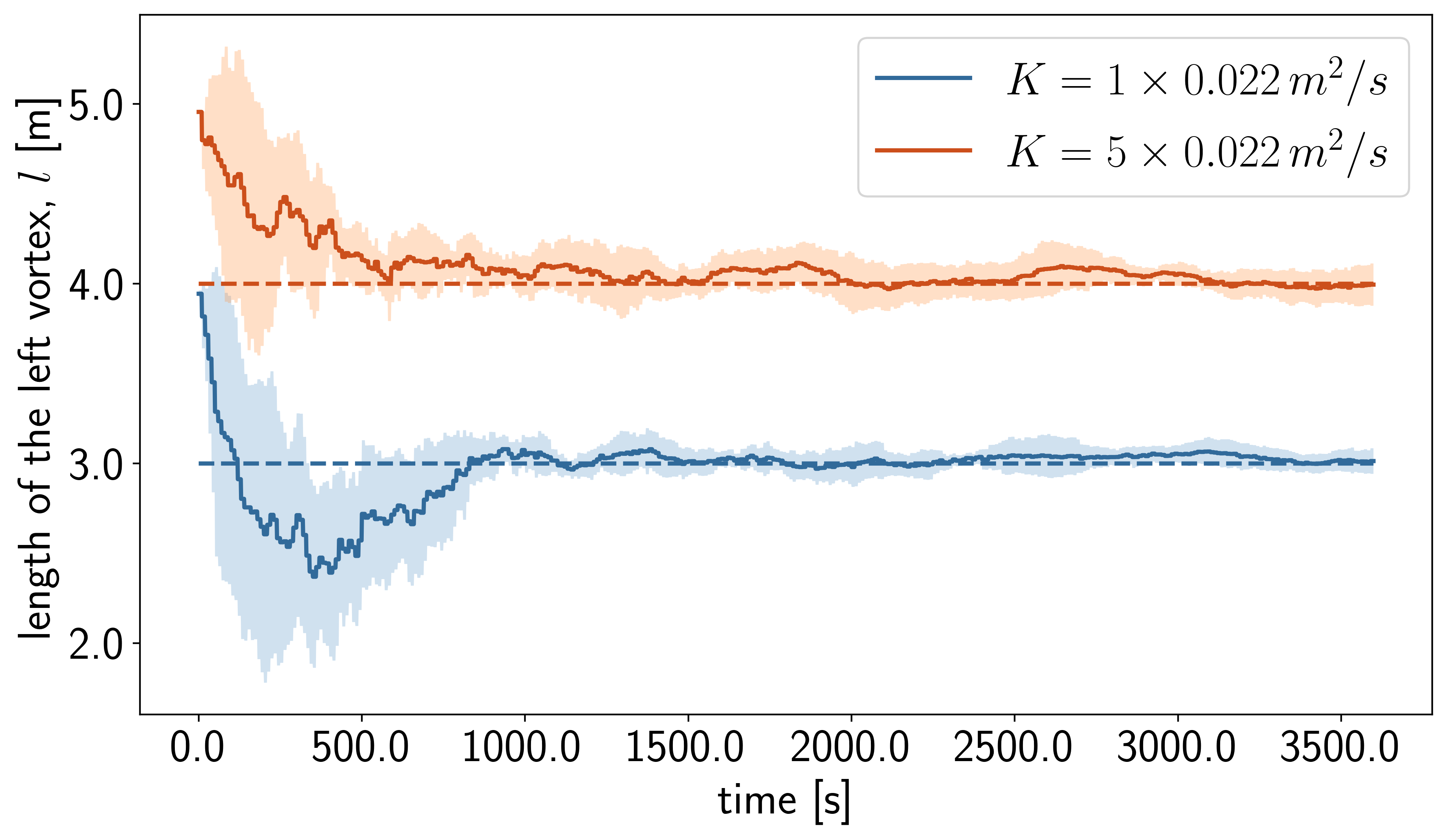}
 \caption{Training performance for learning optimal length of the left vortex ($l$) of the double-vortex field, averaged over 10 independent runs for two values of the diffusion coefficient. The solid lines and the shaded areas show the mean and the variance (one standard deviation) of the learned vortex length, respectively, over the 10 runs. The dashed lines show the respective ground‐truth optimal lengths found via brute-force simulations.}
 \label{Fig:DVortex_PPO}
\end{figure}

Before attempting to solve any of the control problems outlined in section \ref{Section:Control}, we do a mesh study to find an adequately-fine mesh size for the simulations. We use a uniform finite element mesh over the entire domain consisting of cells, which in 2D are triangles with straight sides. The mesh size parameter is the tuple $(n_x, n_y)$, where, $n_x$ and $n_y$ specify the number of rectangles (each divided into a pair of triangles) in the $x$ and $y$ directions, respectively. The total number of triangles (cells) thus becomes $2 \times n_x \times n_y$. For the mesh study we run simulations with parameters in Table \ref{Table:params}, $l= 4.0\,m$, $T=600\,s$, and a range of values for $n_x=2n_y$. Figure \ref{Fig:MeshStudy} depicts the performance metric $J_e$ calculated over the entire room ($\Omega = \text{entire room}$) as a function of $n_x$. As shown in the figure, there is not much improvement in the results for mesh sizes finer than $n_x=80$; thus, we set $n_x=2n_y=80$ for all the subsequent simulations in this study.
For both control problems we assume a center location for the COVID-19 source at $\bm{x}_c=(6,2)\,m$, and the time period of interest as $T = 600 \, s$. Also, parameters in Table \ref{Table:params} are used for all the simulations unless otherwise specified. For the first control problem, i.e. Eq.\ref{Eq:ControlP1}, we first find the ground-truth optimal vortex length $l^{\textrm{opt}}$ by brute-force simulations. For this problem we consider the left quarter of the room to be the region of interest; $\Omega=\{\bm{x}; x\leq 2\,m\}$. Figure \ref{Fig:DVortex_GroundTruth} shows how the performance metric varies as the length of the left vortex changes for two different values of pathogen diffusivity. As shown in the figure, the optimal vortex length depends on the pathogen diffusivity. A naive guess for the optimal vortex length will be $l=2\,m$ in the hope of isolating the left quarter from the rest of the room. However, because of the diffusion phenomenon some of the pathogens will still find their way to the left quarter of the room in spite of the vortices. For this very reason, the left vortex needs to be extended beyond the region of interest, i.e. $l^{\textrm{opt}}>2$. In fact, the more diffusive the pathogens are the longer the left vortex should be. Simulation results in Fig.\ref{Fig:DVortex_GroundTruth} corroborates this; the optimal left vortex length for $K=0.022\, m^2/s$ and $K= 5\times 0.022\, m^2/s$ is $3\,m$ and $4\,m$, respectively.
We use the default parameter values for the PPO algorithm except those in Table \ref{Table:PPOparams} where we provide the parameter values we have used in all our simulations. We use neural networks with ReLU activation functions for both policy and value functions. For the first control problem, we train the control agent for a total time of $6 \times T = 6 \times 600 \, s$. The environment is reset every $T$ seconds. In all the simulations, the time increment is one second, that also means every step from the perspective of the RL agent takes one second. Figure \ref{Fig:DVortex_PPO} shows the training performance of the RL agent in learning the length of the left vortex for two different values of the diffusion coefficient. The figure illustrates the agent's performance in terms of the mean and variance of the learned policy (left vortex length) over 10 independent runs. As shown in the figure, the agent learns a reliable approximate of the optimal policy in roughly $1000\,s \approx 17\, \textrm{mins}$.
\begin{table}
    \begin{center}
        \begin{tabular}{lr}
        \firsthline
        parameter                               & numerical value \\
        \hline
        learning rate                           & 0.005 \\
        number of steps to run per update       & 10  \\
        mini-batch size                         & 10 \\
        number of epochs                        & 10 \\
        discount factor                         & 0.99\\
        \lasthline
        \end{tabular}
        \caption{Parameters description for the PPO algorithm}
        \label{Table:PPOparams}
    \end{center}
\end{table}
Next, we would like our RL agent to learn the optimal policy for the second control problem as stated in Eq.\ref{Eq:ControlP2}. Just like the first control problem, we first find the ground-truth optimal policy, i.e. the optimal center position of the HP source, by brute-force simulations. For the sake of computational simplicity, we fix the center position in the $y$-direction and optimize for the position in the $x$-direction; $\bm{x}_{hp}=(x_{hp},y_{hp})=(x_{hp},3\,m)$. For this problem, we assume a uniform velocity field of $\bm{v}(x,y)=(-0.015, 0)\,m/s$. We also consider the left half of the room to be the region of interest, i.e., $\Omega=\{\bm{x}; x\leq 4\,m\}$.

\begin{figure}
 \centering
 \includegraphics[width = 1 \linewidth]{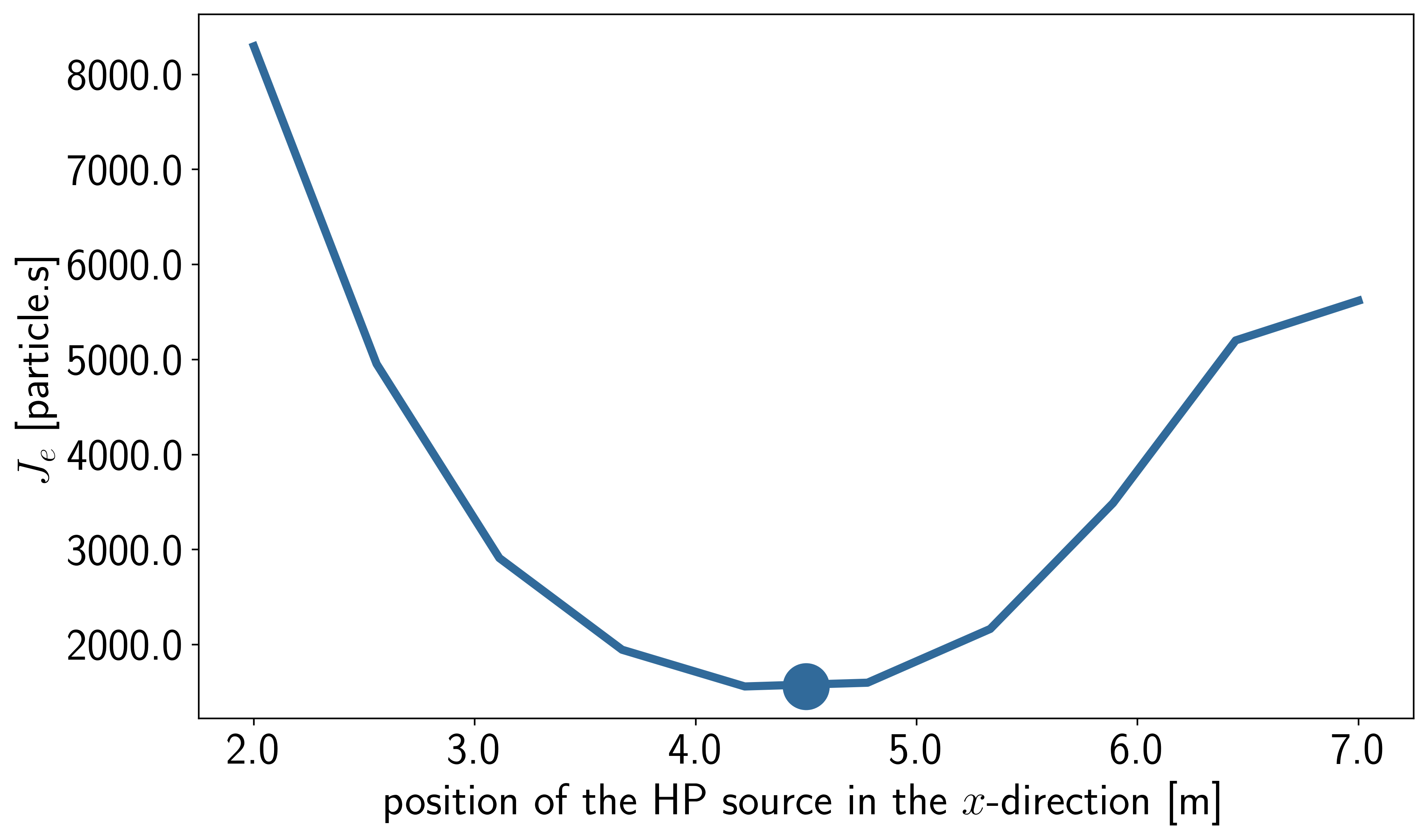}
 \caption{The performance metric, $J_e$ as a function of the HP source location in the $x$ direction. The optimal value minimizing the performance metric is marked by a solid circle.}
 \label{Fig:HP_GroundTruth}
\end{figure}

\begin{figure}
 \centering
 \includegraphics[width = 1 \linewidth]{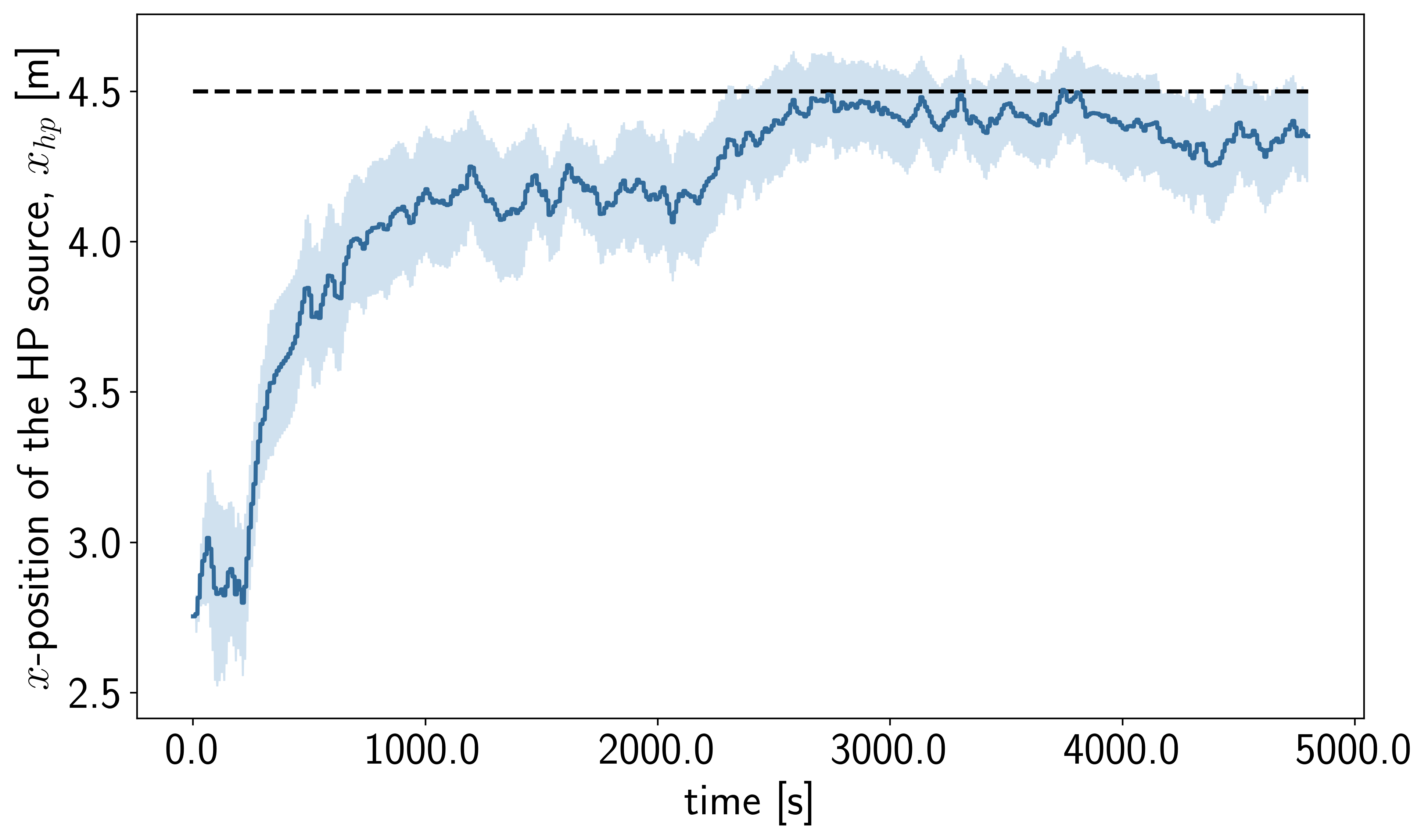}
 \caption{Training performance for learning optimal center position of the HP source in the $x$ direction ($x^{\textrm{opt}}_{hp}$), averaged over 10 independent runs. The solid line and the shaded area show the mean and the variance (one standard deviation) of the learned $x$-position of the HP source, respectively, over the 10 runs. The dashed line show the ground‐truth optimal $x$-position of the HP source.}
 \label{Fig:HP_PPO}
\end{figure}

Figure \ref{Fig:HP_GroundTruth} depicts how the performance measure varies by the $x$ position of the HP source, and that it is minimized at $x^{\textrm{opt}}_{hp}=4.5\,m$. As a first guess, one may go for a position as close to the pathogen source as possible to maximally neutralize the pathogens. However, because of the diffusion, and more importantly, advection in this case, the optimal position for the HP source is not the closest to the COVID-19 source. Figure \ref{Fig:HP_PPO} shows the performance of the RL agent in learning this optimal position over 10 independent runs. All the parameters of the RL algorithm remain the same, except the total simulation time that is extended to $8 \times T = 8 \times 600 \, s$. The simulation results show that the agent learns a reliable approximate of the optimal policy in roughly $3000\,s = 50\, \textrm{mins}$.
\section{Conclusion and future work} \label{Section:Conclusion}
In this study we investigated data-driven control of COVID-19 in indoor environments. We put forward the idea of designing indoor airflow to contain spread of viral pathogens. We formulated the control problem in a general set-up and employed the PPO RL algorithm to learn the optimal control law, i.e. the optimal airflow. Transport dynamics of the pathogens are modeled by advection-diffusion equations and a parameterized double-vortex field is chosen as a class of velocity fields to be optimized by the control agent. Simulation results show that the agent can learn the optimal lengths of the vortices in less than $17 \, \textrm{mins}$.

As a secondary control problem, we also studied the feasibility of optimal placement of disinfectants in a room in order to minimize the infection risk of occupants in a sub-space of the room. We showed that our learning-based controller can learn the optimal location of the disinfectant in less than $50\, \textrm{mins}$. Given the computational complexity of the CFD simulations, lack of knowledge about the fluid-physics of pathogens transport, and frequent change of interior layout of enclosed spaces, the data-driven nature of the proposed ideas makes them particularly advantageous over their model-based counterparts.

Despite its simplicity, the double-vortex velocity field may not be a good choice for designing an effective airflow in more complex built environments, such as, large offices or theaters with many cubicles and seats. In this case, one can employ more complex velocity fields, e.g., multiple-vortex or a number of point vortices, and optimize for their geometric centers and intensities. Also, for the second control problem, more than one disinfectant could be used and optimized for.

Another limitation of the current study is that we looked for time-invariant optimal solutions in both control problems, i.e., a time-invariant optimal double-vortex and a time-invariant optimal location for the disinfectant. This was a good solution mainly because we assumed a stationary source of virus, as well as, a long enough period of interest ($T$) for the pathogens to diffuse after the initial transient. However, if very short transient time is of interest or the pathogen source is moving, a time-varying solution (double-vortex with time-varying length or disinfectant with time-varying location) will probably be much more effective. This will be the future work and an extension to the current study.
%

In general, the proposed data-driven control framework in this study can have significant societal and economic benefits by setting the foundation for an improved methodology in designing case-specific infection control guidelines that can be realized by affordable HVAC devices and disinfectants. Implementing the proposed design and control guidelines helps mitigate the spread of airborne diseases, such as COVID-19, and hence, can save tens of thousands of lives worldwide. In the case of COVID-19 or other potential pandemic-causing viruses, containing the indoor virus spread will also facilitate reopening of schools, universities, offices, and restaurants, to name a few, which in turn, speeds up the recovery of the US and the world economy. This will help low-income communities in particular, by allowing small businesses to open up their operation to avoid major income loss in these vulnerable communities.
\bibliography{references}

\newpage
\appendix

\section{Variational (weak) formulation}

In this section we present the variational form of the PDEs in Eqs.\ref{Eq:ControlP1} and \ref{Eq:ControlP2} required for the FEniCSx finite-element solver. A straightforward approach to solving time-dependent PDEs by the finite element method is to first discretize the time derivative by a finite difference approximation, which yields a sequence of stationary problems, and then turn each stationary problem into a variational formulation. We use backward Euler, aka implicit Euler discretization:

\begin{equation} \label{Eq:Euler}
  \left(\frac{\partial c}{\partial t}\right)^{n+1} = \frac{c^{n+1}-c^n}{\Delta t},    
\end{equation}
where, superscript $n$ denotes the quantity at time $t_n$ and $\Delta t$ is the time discretization parameter.

The basic recipe for turning a PDE into a variational problem is to multiply the PDE by a function $w$, integrate the resulting equation over the domain $D$, and perform integration by parts of terms with second-order derivatives. The function $w$ which multiplies the PDE is called a \textit{test} function. The unknown function $c$ to be approximated is referred to as a \textit{trial} function. The trial and test functions belong to certain so-called \textit{function spaces} that specify the properties of the functions. An important feature of variational formulations is that the test function $w$ is required to vanish on the parts of the boundary where the solution $c$ is known. Now multiplying the advection-diffusion equation in Eq.\ref{Eq:ControlP1} by the test function $w$, integrating it over the domain $D$, performing integration by parts, and applying the test function properties and boundary conditions, we arrive at the variational form:
\begin{align} \label{Eq:Weak1}
& \int_D \left( \frac{1}{\Delta t} \left(c^{n+1}-c^n\right)w+\left(\bm{v}.\nabla c^{n+1}\right)w+K\nabla c^{n+1}.\nabla w\right)\, d\bm{x} \nonumber\\ 
-&\int_D f^{n+1} w\, d\bm{x} +\int_D \lambda c^{n+1} w \, d\bm{x} = 0.
\end{align}

The variational problem now is to find $c$ from the trial space such that Eq.\ref{Eq:Weak1} holds true for all the test functions, $w$, in the test space. This variational problem is a \textit{continuous problem}: it defines the solution $c$ in the infinite-dimensional function space (the trial space). The finite element method finds an approximate solution of the continuous variational problem by replacing the infinite-dimensional function spaces by discrete (finite-dimensional) trial and test spaces. FEniCS automatically solves the discrete variational problem.

With the same procedure, we derive the variational form of the coupled advection-diffusion equations in Eq.\ref{Eq:ControlP2} as:

\begin{align} \label{Eq:Weak2}
& \begin{aligned}
\int_D \biggl( \frac{1}{\Delta t} \left(c^{n+1}-c^n\right)w_1&+\left(\bm{v}.\nabla c^{n+1}\right)w_1\\
&+K\nabla c^{n+1}.\nabla w_1\biggr) d\bm{x}
\end{aligned}\nonumber\\
&\begin{aligned}
+\int_D \biggl( \frac{1}{\Delta t} \left(c^{n+1}_{hp}-c^n_{hp}\right)w_2&+\left(\bm{v}.\nabla c^{n+1}_{hp}\right)w_2\\
&+K_{hp}\nabla c^{n+1}_{hp}.\nabla w_2\biggr) d\bm{x}
\end{aligned}\nonumber\\ 
- & \int_D \left(f^{n+1} w_1+f^{n+1}_{hp} w_2\right)\, d\bm{x} \nonumber\\
+&\int_D \left(\lambda c^{n+1} w_1+\lambda_{hp} c^{n+1}_{hp} w_2\right)  d\bm{x} \nonumber\\
+ & \int_D \left( \alpha_1 c^{n+1} c^{n+1}_{hp} w_1 + \alpha_2 c^{n+1} c^{n+1}_{hp} w_2\right) d\bm{x}= 0,
\end{align}

%
where, $w_1$ and $w_2$ are test functions for the unknown variables $c$ and $c_{hp}$, respectively.
%
\begin{IEEEbiography}[{\includegraphics
[width=1in,height=1.25in,clip,
keepaspectratio]{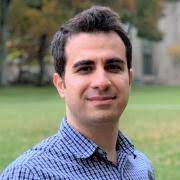}}]
{Ashkan Haji Hosseinloo} received the B.Sc. from Amirkabir University of Technology, Tehran, Iran, in 2009, the M.Eng. from Nanyang Technological University, Singapore, in 2013, and the Ph.D. from the Massachusetts Institute of Technology, Cambridge, MA, USA, in 2018, all in Mechanical Engineering. Ashkan is a postdoctoral scholar at MIT Laboratory for Information and Decision
Systems (LIDS) and MIT Institute for Data, Systems, and Society (IDSS). His research lies at the intersection of machine learning and system
dynamics \& control, and is motivated by the urgent need to address the pressing issues of energy and environmental sustainability and social equity. Ashkan’s work has addressed fundamental challenges in the control of complex dynamical systems with applications in structural dynamics, energy harvesting, and smart cities.
\end{IEEEbiography}
\begin{IEEEbiography}[{\includegraphics
[width=1in,height=1.25in,clip,
keepaspectratio]{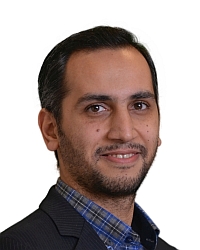}}]
{Saleh Nabi} received the B.Sc. from K. N. Toosi University of Technology, Tehran, Iran, in 2005, the M.Sc. from Isfahan University of Technology, Isfahan, Iran, in 2008, and the Ph.D. from University of Alberta, Canada, in 2013, all in Mechanical Engineering. Saleh is a Principal Research Scientist at Mitsubishi Electric Research Labs (MERL). His research interests are at the intersection of fluid mechanics, scientific machine learning, dynamical systems, and optimal control in complex systems. His current research involves hybrid methods using traditional tools along with deep learning-based methods for efficient and robust control and estimation of PDEs with applications to HVACs and atmospheric LiDARs.
\end{IEEEbiography}
\begin{IEEEbiography}[{\includegraphics
[width=1in,height=1.25in,clip,
keepaspectratio]{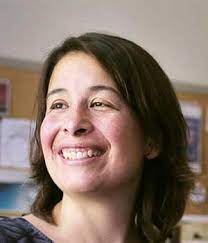}}]
{Anette (Peko) Hosoi} received the B.A. degree from Princeton University, Princeton, NJ, USA, in 1992, and the M.Sc. and the Ph.D. degrees from the University of Chicago, Chicago, IL, USA, in 1994 and 1997, respectively, and all three degrees in physics. She is the Neil and Jane Pappalardo Professor of Mechanical Engineering and associate dean of engineering at the Massachusetts Institute of Technology, Cambridge, MA, USA. She is the co-founder of the MIT Sports Lab which connects the MIT community with pro-teams and industry partners to address data and engineering challenges in the sports domain. Hosoi's research interests include fluid dynamics, unconventional robotics, and bio-inspired design. She has received numerous awards including the APS Stanley Corrsin Award,  the Bose Award for Excellence in Teaching, and the Jacob P. Den Hartog Distinguished Educator Award.
\end{IEEEbiography}
\vspace{-4in}
\begin{IEEEbiography}[{\includegraphics
[width=1in,height=1.25in,clip,
keepaspectratio]{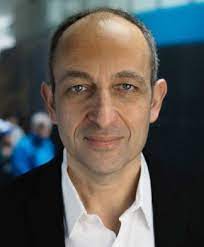}}]
{Munther A. Dahleh} received the B.S. degree from Texas A\&M University, College Station, TX, USA, in 1983, and the Ph.D. degree from Rice University, Houston, TX, USA, in 1987, both in electrical engineering. He is the William A. Coolidge Professor with the Massachusetts Institute of Technology, Cambridge, MA, USA,
where he is also the Director of the Institute for  Data, Systems and Society.
Dahleh is a co-recipient of four George S. Axelby Outstanding Paper Awards. He is internationally known for his fundamental contributions to robust control
theory, computational methods for controller design, the interplay between information and control, the fundamental limits of learning and
decision in networked systems, and the detection and mitigation of systemic risk in interconnected and networked systems.
\end{IEEEbiography}

\end{document}